# Arbitration Among Vertical Services

Claudio Casetti, Carla Fabiana Chiasserini
Politecnico di Torino, Italy

Thomas Deiß*
Nokia Solutions and Networks, Germany

Giada Landi
Nextworks, Italy

Nuria Molner, Jorge Martín-Pérez
IMDEA Networks Institute and
Universidad Carlos III de Madrid, Spain

Cao-Thanh Phan, Farouk Messaoudi
BCOM, France

Juan Brenes Baranzano
ATOS, Spain

*Abstract*—A 5G network provides several service types, tailored to specific needs such as high bandwidth or low latency. On top of these communication services, verticals are enabled to deploy their own vertical services. These vertical service instances compete for the resources of the underlying common infrastructure. We present a resource arbitration approach that allows to handle such resource conflicts on a high level and to provide guidance to lower-level orchestration components.

## I. INTRODUCTION

5G networks will offer a high degree of support for the operational requirements of vertical industries, such as automotive, media and entertainment, and e-health. 5G networks can tailor the infrastructure to satisfy higher-layer requirements, leveraging the softwarization and virtualization of the infrastructure. E.g., for ultra-reliable low-latency services the User Plane Functions (UPF) may be deployed close to the mobile edge, whereas for enhanced Mobile Broadband the UPFs may be centralized to increase multiplexing gains. Verticals from different industries are enabled to define and deploy their vertical services on top of the 5G networks.

The 5G-TRANSFORMER system enables verticals to define their services in an easy manner [1]. Verticals may select a Vertical Service Blueprint (VSB) from a catalogue, complete it, and eventually instantiate the service. The orchestration of the infrastructure is handled by the system, the verticals do not have to care about it.

As new services are instantiated and the amount of resources in the infrastructure varies over time, some Vertical Service Instances (VSI) might get less resources than required to satisfy their Service Level Agreements (SLA). In this paper, we present an approach to arbitrate resources among VSIs: Firstly, a vertical agrees on a resource budget with the provider, e.g. for storage, bandwidth, processing, and it assigns priorities to its VSIs. Secondly, an Arbitrator component within the system assigns resources from the resource budget to the VSIs based on priorities. For some resources, e.g.

This work has been partially funded by the EU H2020 5G-Transformer Project (grant no. 761536). * Corresponding author email: thomas.deiss@nokia.com

processing and bandwidth, the needed resources depend on placement decisions done by lower layers. In this case, the Arbitrator determines several Deployment Flavours (DF) for different placement options. The DFs are used by the lower layers in the system for the actual deployment decisions.

The Arbitrator handles also the mapping of VSIs to Network Slice Instances (NSI), again mapping resources and calculating DFs. The Arbitrator allows to keep the business logic about priorities among vertical services and agreed resource budgets in the high-level components, still allowing the lower-level components to make placement and scaling decisions on their own.

We firstly describe the 5G-TRANSFORMER system and the component interacting with the vertical, see Section II. Then we describe the Arbitrator component and its roles, see Section III. In Section IV we present specific resource assignment algorithms. We summarize the paper in Section V.

## II. 5G-TRANSFORMER ARCHITECTURE

The 5G-TRANSFORMER architecture [1], [2], presented in Figure 1, is a three-layer architecture that relies on the Network Function Virtualization (NFV) principles of physical infrastructure virtualization, management and orchestration of virtual resources, and dynamic instantiation of services.

At the bottom, the Mobile Transport and Computing Platform (5GT-MTP) [3] manages a physical infrastructure composed of multi-technology network resources, spanning from the radio to the (possibly integrated) fronthaul and backhaul, access and transport domains, interconnecting computing and storage resources placed in centralized or in distributed Multi-Access Edge Computing (MEC) datacenters.

Infrastructure resources are virtualized and abstracted by the 5G-MTP to allow the instantiation of Virtual Network Functions (VNFs) and NFV Network Services (NFV-NSs), orchestrated by the Service Orchestrator (5GT-SO) [4]. The 5GT-SO implements the functionalities of an NFV Orchestrator (NFVO), enhanced with management of MEC resources and inter-domain federation. The federation enables the dynamic composition and provisioning of end-to-end NFV-NS, deployed across multiple administrative domains.

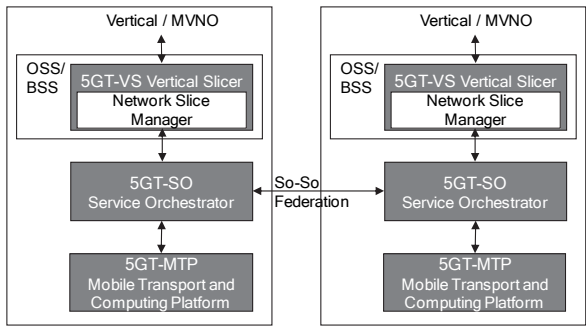

*Figure 1: 5G-TRANSFORMER Architecture*

On top of the 5GT-SO, the Vertical Slicer (5GT-VS) [5] offers a vertical-oriented perspective, where vertical services are requested and deployed based on business requirements (e.g. in terms of required functionalities, service constraints and parameters), delegating the infrastructure and resource related decisions to the lower layers. The 5GT-VS translates a service specification into the corresponding resource requirements and instantiates network slices satisfying the requirements, complying with the SLAs of the vertical. The management of network slices to deliver vertical services is the key to handle business level constraints like service priority, resource budgets, requirements for service isolation and sharing, or composition of specific service functions from trusted operators, 3rd party service providers or verticals. The 5GT-VS maps each vertical service request into NFV-NS Descriptors (NSD) capturing the infrastructure and resource-based characteristics of the service and, through the Arbitrator component, it takes decisions about new or existing NSIs to be instantiated, re-used or modified to serve the request and deploy the service.

### III. ARBITRATOR

We shortly describe the context of the Arbitrator, i.e. the internal architecture of the 5GT-VS. Thereafter we describe its roles within the system.

*A. Vertical Slicer Design*

The internal architecture of the 5GT-VS [6] is shown in Figure 2. The core of the 5GT-VS are three main components providing its procedural and algorithmic logic. Additional components provide administrative functions (e.g. management of tenants or SLAs), catalogues for blueprints and descriptors, records of VSIs and Network Slice Instances (NSI), management of external interfaces to verticals and 5GT-SO and monitoring features. The three main components are:

(i) the *VSI/NSI coordinator and lifecycle manager* handles the delivery and runtime of vertical services, coordinating their lifecycle with suitable actions at the corresponding NSI and network slice subnet instances (NSSI). Actions on NSIs are performed requesting the instantiation, scaling, modification or termination of the related NFV NS instances at the underlying 5GT-SO. They are triggered based on verticals' requests (e.g. to instantiate a new service) or in reaction to monitoring notifications signaling a breach in the SLA;

(ii) the *VSD/NSD translator* handles the translation between vertical service descriptors (VSD) and NSDs, allowing to shift from the business oriented perspective of the verticals to the resource oriented internal view of the system to deploy the services. The translation is modelled using rules that take as input the VSB and potential ranges of values for the service parameters that can be configured by the verticals in the VSD. As output, the translator selects one or more NSDs with potential DFs defining a network service able to meet the service requirements expressed in the VSD.

(iii) the *Arbitrator* regulates how multiple vertical services get access to a vertical's resource budget and takes decisions about how services are mapped to isolated or shared network slices. Note, the Arbitrator does not have a complete view of the resources, it has to work with limited information of the resources that are available in the infrastructure. It is aware of the SLAs among the verticals and it can balance the resources assigning probabilities to the services to be deployed based on the SLAs of the verticals. Eventually, deployment decisions are made by the 5GT-SO.

Moreover, when resources are overused and the services could not be placed in the network, the Arbitrator reassigns the resources among the verticals based on their SLAs.

In summary, the main tasks of the Arbitrator [6] are:
1. Decide how to map new vertical services in NSIs, allowing multiple vertical services to share one or more NSIs or NSSIs.
2. Determine the DFs of each service, meeting the vertical's Quality of Service (QoS) requirements, while accounting for the services' priority level.

Note, upon the request of a new service instance by a vertical, the Arbitrator may need to update the DFs of previously allocated VSIs.

We expect the 5GT-SO to free resources and to try alternative placements for other VSIs autonomously. To free resources, the 5GT-SO can use resource scaling [7], adapting virtualized resource consumption according to the VNFs' loads. VNF resource scaling is non-service disruptive. In horizontal mode (i.e., scale in/out), the scaling varies the use of resources allocated to a VNF through adding or removing VNF components . Resource scaling can be triggered by an Arbitrator operation or by the 5GT-SO itself. Arbitrator operations to instantiate or terminate a VSI in a shared NSIs or when modifying DFs may trigger resource scaling by the 5GT-SO. The 5GT-SO may trigger resource scaling on its own to adapt resource usage in general to VNF load as well as to control failures and system upgrades and return a VSI to a stable and desired state.

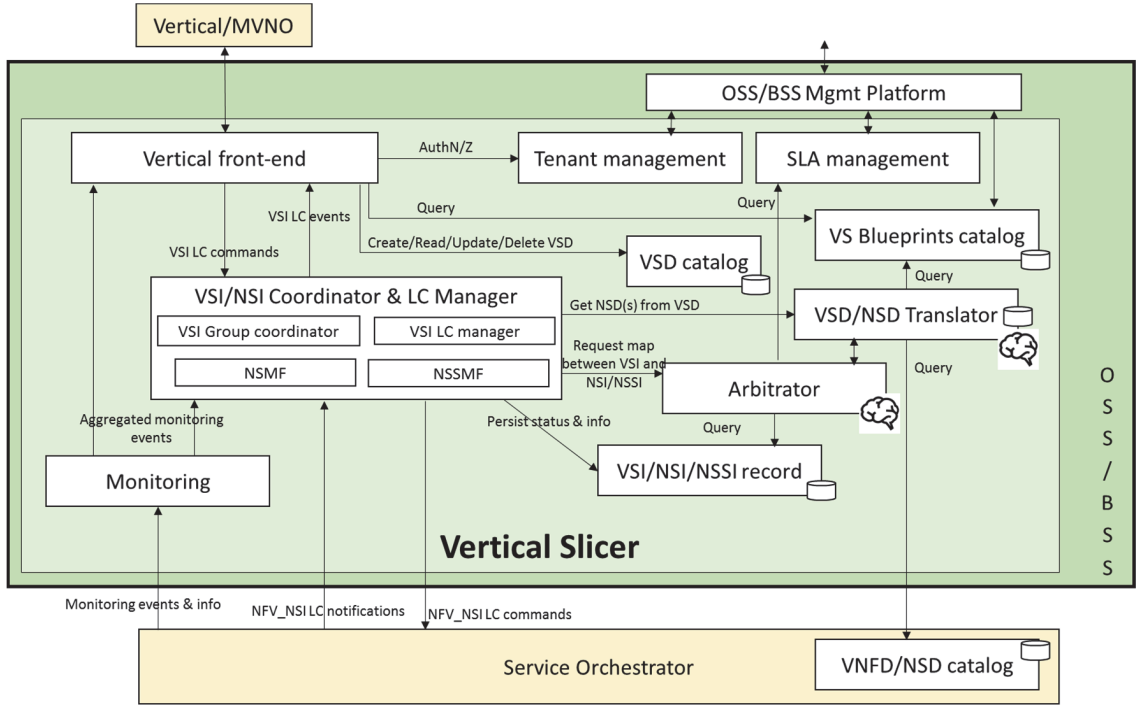

*Figure 2: 5G-TRANSFORMER Vertical Slicer architecture*

*B. Interaction with Placement*

The amount of needed resources for a VSI may depend on placement decisions for VNFs and virtual links, e.g. different amounts of bandwidths or processing resources may be needed. When VNFs are deployed in different data centers, more processing resources might speedup processing and compensate as far as possible for transmission delays. But the Arbitrator, and the 5GT-VS in general, do not have information about the specific location of data centers and implied latencies among them. Therefore, it cannot make the placement decision, which are delegated to the 5GT-SO. Thus, the Arbitrator calculates several DFs for different placement option and passes them to the 5GT-SO. The 5GT-SO takes these DFs into account when placing VNFs.

Even with several DFs, resource arbitration for a new VSI may fail out of several reasons:

1) The new VSI cannot be accommodated within the available resource budget. In this case the vertical may choose among a) cancel the instantiation request, b) increase its resource budget from the operator and repeat the instantiation request, or c) confirm the instantiation request without increasing the resource budget. In case c), the Arbitrator reassigns resources from the budget, taking resources from already existing VSIs with lower priority and providing them to the new VSI. Note, this might cause performance degradations for the lower priority VSIs or even termination of these VSIs.

2) There are sufficient resources for the new VSI in the budget, but the 5GT-SO cannot find a placement of VNFs satisfying all requirements. E.g., there are no more compute resources in an edge data center, which would be required to satisfy latency constraints., or a physical link does not have sufficient residual bandwidth for the traffic of the new VSI. If the 5GT-SO cannot free sufficient resources by resource scaling, it informs the 5GT-VS about the resource shortage. The 5GT-VS has a very abstract view of resources only, therefore it cannot know what resource exactly is missing. But it can know which VNF or Virtual Link (VL) cannot be provided the needed resources and it can inform the vertical accordingly. The vertical can decide then to undo the vertical service instantiation, to still deploy the VSI with insufficient resources, or to deploy it only in those areas with sufficient resources. Which of these options is suitable depends on the specific vertical service and has to be decided by the vertical.

For sake of simplicity we do not differentiate among resources of the same type at different locations, e.g. compute resources in a central cloud data center are not distinguished from those in a local office or even at the mobile edge. Nevertheless, the approach presented here can be generalized to provide such differentiation. Firstly, a vertical has to agree separate resource budgets with the provider for these resource types. Secondly, the Arbitrator has to calculate additional DFs using such different resources. The resource assignment algorithms described subsequently focus on the simpler case.

IV. RESOURCE ASSIGNMENT

Resources are assigned to verticals by the Arbitrator based on their service requests. For each VSI of a vertical, the Arbitrator determines the associated DFs, such that the vertical's QoS requirements are met and accounting for the service priority level. The assignment of resources requires agreeing between the parties (vertical and slice provider, i.e., the 5GT-SO) on service objectives, needs and available resources. The vertical specifies its needs in terms of Service Level Objectives (SLOs). The slice provider, instead, will define service level classes and guarantees on resource availability through Service Level Agreement Templates (SLAT). The matching of SLOs desired by the vertical and SLATs offered by a provider will converge into an SLA.

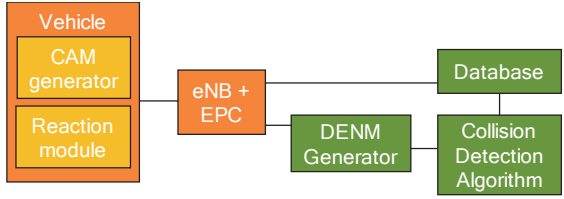

*Figure 3: VNFFG of the ICA service*

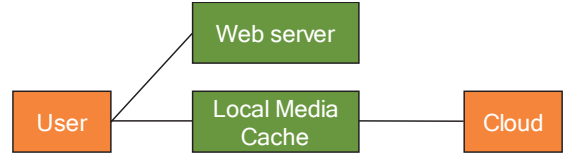

*Figure 4: VNFFG of the MS service*

The Arbitrator assigns resources among VSIs of the same vertical. As an example, we consider a vertical from the automotive domain, whose SLA foresees the availability of CPU (C), bandwidth (B), memory (M), and storage (S). In this example, the vertical wants to deploy two VSIs: an Intersection Collision Avoidance (ICA) service and a Multimedia Streaming (MS) service, with ICA having priority over MS.

At first, the Arbitrator considers the highest-priority service instance, ICA in our case, and allocates memory and storage based on the needs exhibited by the VNFs in the VNF Forwarding Graph (VNFFG) representing it (see Figure 3). CPU and bandwidth allocation, instead, is more complex, as it depends on the VNF placement decisions by the 5GT-SO. Let us assume that the main performance metric of service $s$ is the maximum latency $D_s$, , which depends on two components. Firstly, it depends on the processing time to execute the VNFs, given by the CPU allocated to the VNFs execution. Secondly, if the 5GT-SO chooses different servers for the implementation of the VNFs, it depends on the network travel time, due to the time needed to transfer data from one VNF to the next in the VNFFG, and on the bandwidth associated with the VL connecting the servers. Specifically, for a set of VNFs $V$, the Arbitrator will compute the allocation for the ICA service in terms of the CPU, $\mu$ and bandwidth, $\beta$ that satisfy the following conditions:

$$\sum_{v \in V} \frac{1}{f_v \mu - \lambda_v} + \sum_{(u,v) \in E} \frac{d_{u,v}}{f_{u,v} \beta} \leq D_s \quad (1)$$

$$\mu \leq C \;;\; \beta \leq B \;;\; \mu/\beta = C/B \quad (2)$$

where $f_v$ is the relative computational requirement of VNF $v \in V$, $f_{u,v}$ is the relative bandwidth requirement for the VL connecting VNFs $u,v \in V$ and $d_{u,v}$ is the amount of data exchanged by $u$ and $v$. (1) accounts for the latency due to both the VNF execution (modelling the generic VNF $v$ as a FIFO M/M/1 queue with $f_v \mu$ as the output rate and $\lambda_v$ as the service request rate input to $v$) and the travel time over the VLs connecting any two adjacent VNFs; (2) imposes that both the total CPU and bandwidth allocations do not exceed the corresponding budget available to the vertical; it also imposes that the ratio between $\mu$ and $\beta$ be equal to the ratio of $C$ and $B$ in order to ensure a consumption of the different types of resources proportional to the corresponding budgets. Finally, it is worth pointing out that we neglected the delay due to memory/storage access.

Conditions (1) and (2) represent a *worst-case* scenario, but if all VNFs are deployed on the same server, the dependence on network travel time can be neglected, leading to a *best-case* scenario. Thus, the bandwidth required for data transfer can be set to zero and the allocated CPU can be computed as:

$$\sum_{v \in V} \frac{1}{f_v \mu - \lambda_v} \leq D_s \quad (3)$$

$$\mu \leq C \quad (4)$$

The Arbitrator will then proceed with the MS service (whose VNFFG is shown in Figure 4), following the same steps as above but using the remaining budget available to the vertical in terms of CPU and bandwidth. The MS uses only two virtual applications: one that caches local copies of the video files, and effectively serves the video files to the user; the other one is a web server providing the video player to the user through Javascript.

Let us now provide a numerical example. We start by listing the SLOs:

ICA Service Level Objectives:
- Geographical coverage radius: 500 m
- Maximum latency $D_{ICA}$: 20ms
- Maximum service request rate: 60/s

MS Service Level Objectives:
- Geographical coverage radius: 1000 m
- Maximum latency $D_{MS}$: 5s
- Minimum data rate: 200 kb/s
- Maximum number of simultaneous streams: 6

We will also assume (considering experimental results from consumer-grade hardware) C=10000 packets/s and B=10Gbit/s and that the relative computational requirement $f_v$ for the database and VNFs of the ICA service are both 0.05, which yields $f_v$ for the collision detector equal to 0.9. For the MS service, we assume $f_v$ = 0.95 for the media cache and 0.05 for the Web server.

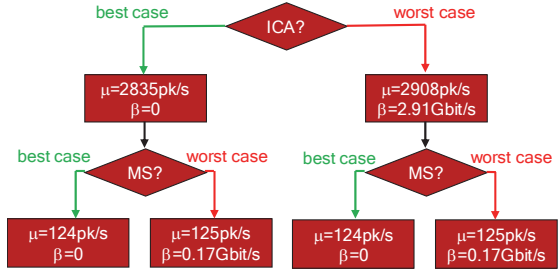

*Figure 5: Best- and worst-case allocations for ICA and MS services*

The solutions of equations (1)-(4) using numerical methods for both the worst and the best case can be summarized by the tree structure of Figure 5. Note that the 5GT-VS will send to the 5GT-SO within DFs the minimum and maximum values of required CPU and bandwidth for ICA and MS, namely, $\mu_{ICA} \in [2835, 2908]$ packets/s and $\beta_{ICA} \in [0, 2.91]$ Gbit/s and $\mu_{MS} \in [124, 125]$ packets/s and $\beta_{MS} \in [0, 0.17]$ Gbit/s.

For services of several verticals, the Arbitrator decides on the deployment of services based on the SLAs, starting with the verticals that have highest priority in the agreements. Verticals of the same priority have the same probability to be deployed:

$$P(v) = \frac{1}{|I_v|} \quad (5)$$

where *P(v)* is the probability of selecting one vertical of this priority, and $I_v$ is the set of verticals in the priority considered. This guarantees that all the verticals in the same priority have the same probability to be chosen to start the deployment. In addition, we remove the ones that are considered each time from the set to continue with the ones that have not been deployed already. This guarantees that a vertical is not selected several times while another one is not selected at all, which makes the probability non-dependent on the number of services of a vertical.

This algorithm can be extended to the case of VNFs shared among different VSIs, even of different verticals. In this case, the mapping of VSIs to NSIs and resource assignment is intertwined. If a vertical requests to instantiate a vertical service using VNFs already deployed for some other VSI, the Arbitrator will check SLAs for isolation requirements. If the new or the existing VSIs are required to be isolated, the Arbitrator cannot share these VNFs and will map the newly requested vertical service to a new NSI and the previous algorithm can be used. If there is no requirement to isolate the new and the existing VSIs, the Arbitrator can map the newly requested service to an existing NSI and share the VNFs. The Arbitrator will modify the DFs of shared VNFs increasing the amount of resources, such that the VNFs can handle the traffic of the new VSI as well. This is translated in the formulae by modifying the arrival and processing rates and comparing them against the minimum of the latency requirement of the VSIs.

$$\sum_{v \in V} \frac{1}{\sum_{s \in S}(\mu_v^s - \lambda_v^s)} + \sum_{(u,v) \in E} \frac{d_{u,v}}{\beta_v} \leq D_s \quad (6)$$

The arrival and processing rates now depend on all the services that use the same instance of the VNF. *S* is the set of services that share at least one VNF instance with the demand we are placing now and it appears in the first part of the formula modifying the total delay.

An initial implementation of the 5GT-VS prototype shows that the overhead added to the total provisioning time for an end to end service is minimal, in the order of 1%. For example, considering the provisioning of a simple virtual Content Delivery Network with one origin server and two caches, all deployed as VNFs, the 5GT-VS processing is in the order of 1.8 seconds, while the provisioning time at the NFVO and Virtual Infrastructure Manager level takes more than 2.5 minutes, with most time needed for the creation of the virtual machines and the configuration of the applications.

## V. CONCLUSION

In this paper we described the Arbitrator component, which handles both the mapping of vertical services to network slices as well as resource assignments to vertical services, thereby taking care of priorities among the vertical services. We have shown how different deployment flavours can be determined such that resource assignments can be made in the vertical slicer, while still leaving the actual placement decisions to the underlying service orchestrator. We described the basic algorithm for resource assignments among vertical services of one vertical and showed how to extend it assign resources to services of several verticals and to the case of shared VNFs.